# A family of greedy algorithms for finding maximum independent sets

Asbjørn Brændeland

*Abstract*

The greedy algorithm $A$ iterates over a set of uniformly sized independent sets of a given graph $G$ and checks for each set $S$ which non-neighbor of $S$, if any, is best suited to be added to $S$, until no more suitable non-neighbors are found for any of the sets. The algorithms receives as arguments the heuristic $h$ used to evaluate the independent set candidates, and the initial cardinality $k$ of the independent sets. In the most difficult cases, with the simplest heuristics and $k = 1$, $A$ returns a correct result every time, for $n \leq 16$, and 84% of the time, for $n = 100$, when $n = |V(G)|$ With a more sofisticated heuristic and $k = 2$, the succes rate remains a 100% throughout the range. In cases of failure for $n \leq 100$, $\alpha(G) - A(G)$ is hardly ever greater than 1.

*Algorithm*

A greedy algorithm will for each iteration or recursion make a locally optimal choice, in the hope of ending up with a global optimum [1]. Greedy algorithms for finding maximum independent sets, in general, or targeted on graphs with various characteristica, have been studied for several years. This paper simply presents some test results for a family of greedy general MIS algorithms.

Algorithm $A$, which is a modification and extension of a part of the algorithm given in [4], searches for a maximum independent set in a given graph $G$ as follows:

1. Let $S$ be a the set of independent sets of a given cardinality in $G$.
2. Let $T = \emptyset$.
3. For each set $S_i$ in $S$,
    3.1. let $U \subset V(G)$ be the set of non-neighbors of $S_i$, and,
    3.2. if $U \neq \emptyset$,
        3.2.1. evaluate each vertex in $U$ as a candidate to join $S_i$,
        3.2.2. select the vertex $s$ that got the highest score and
        3.2.3. add $S_i \cup \{s\}$ to $T$.
4. If $T = \emptyset$, return $S$,
5. else, set $S = T$, and go to 2.

An implementation of the algorithm must receive two arguments $h$ and $k$, in addition to $G$:

- $h$ is the heuristic used to evaluate the non-neighbors of $S_i$ as MIS candidates, and
- $k$ is the cardinality of the initial elements in $S$.

The non-neighbor best suited to be added to an expanding independent set, is

    **a**. the one with the most non-neighbors of its own,

or

    **b**. the one that, together with its own non-neighbors, induces the most stable graph,

when the stability of a graph $H \subset G$ on $o$ vertices in this context is given by the formula

$$\sum_{v \in V(H)} \frac{o}{\deg_H(v) + 1} \quad (1)$$

$S$ can initially contain singletons, pairs, or larger tuplets—in principle of cardinalities up to the independence number of $G$, $\alpha(G)$.

$A_{hk}$ then denotes an instance of $A$ that receives the heuristic $h \in \{\mathbf{a}, \mathbf{b}\}$ and the initial independent set cardinality $k$. An instance of $A$ is an algorithm in the family $\mathcal{A}$.

*Testing*

The tests were done on random graphs, when in a random graph $(V, E)$ of a given order and size, the members of $E$ have been selected at random from the handshake product of $V$.

Test 1.

Algorithms $A_{a1}$, $A_{b1}$, $A_{a2}$ and $A_{b2}$ were tested, with the USBE-tree search described in [2] as control, on graphs on $n$ vertices and $m$ egdes, with $n$ ranging from 12 to 42 and $m$ ranging from $2n$ to a little less than $\binom{n}{2}$ (the the number of graphs on $n$ vertices approaches 1 when $m$ approaches $\binom{n}{2}$) with 100 runs for each pair $(n, m)$. This gave about one million test runs.

The sizes that gave the highest failure rates in Test 1. lay in the range $3n$ to $5n$, so in Test 2. and 3. $m$ was always set to $4n$.

Having observed in Test 1. that not more than two out of four algorithms failed with the same input, the chances of triple and quadruple failures were deemed to be negligible within any testable range, so in some of the subsequent tests three or four of the greedy algorithms were tested without additional control, using the maximum of the returned cardinalities in each run as measure.

Test 2.

Algorithms $A_{a1}$, $A_{b1}$, $A_{a2}$ and $A_{b2}$ were tested with regard to failure rates, for $n$ ranging from 20 to 100, starting with 300 000 test runs for $n = 20$ and ending with 580 test runs for $n = 100$. The reasons for the decrease in number of runs were that the lower the failure rates the more runs were required to get consistent results, and, most importantly, the run time required by $A_{b2}$ (see *Complexity*).

Test 3.

Algorithms $A_{a1}$, $A_{b1}$ and $A_{a2}$ were tested with regard to accuracy in 60 000 test runs, 1000 for each $n$, ranging from 20 to 80.

*Failure ratios*

In Test 1. the failure ratio for $A_{a1}$ went from zero for $n = \leq 16$, to a little less than 0.004 for $n = 42$. For $A_{b1}$, 9 failures occured in the entire range, and for the others, none.

The results from Test 2 are given in the table below.

| n | m | runs | $A_{a1}$ | $A_{b1}$ | $A_{a2}$ | $A_{b2}$ |
|---|---|---|---|---|---|---|
| 20 | 80 | 300,000 | 0.00019 | 0 | 0 | 0 |
| 30 | 120 | 200,000 | 0.00230 | 0.00001 | 0 | 0 |
| 40 | 160 | 100,000 | 0.00947 | 0.00003 | 0 | 0 |
| 50 | 200 | 100,000 | 0.02064 | 0.00024 | 0.00003 | 0 |
| 60 | 240 | 50,000 | 0.03674 | 0.00064 | 0.00004 | 0 |
| 70 | 280 | 10,000 | 0.06130 | 0.00050 | 0.00003 | 0 |
| 80 | 320 | 5,000 | 0.09320 | 0.00360 | 0.00080 | 0 |
| 90 | 360 | 2,085 | 0,11799 | 0.00480 | 0.00144 | 0 |
| 100 | 400 | 580 | 0.16035 | 0.01207 | 0.00517 | 0 |

Table 1. Failure ratios for 4 members of $\mathcal{A}$.

*Accuracy*

All through Test 1. $\alpha(G) - A(G) \leq 1$. In Test 3. $\alpha(G) - A_{a1}(G) = 2$, once for each of $n = 63, 75$ and $76$.

*Complexity*

By means of the algorithm described in [3] it takes $3\binom{n}{k}$ time to set up the initial list of vertex sets, before the non-independent sets are filtered out.

$A_{\mathbf{a}k}$ starts with $\binom{n}{k}$ $k$-tuplets in $S$, and for each $S_i \in S$, each vertex in $\overline{N}(S_i)$ is evaluated. The total number of evaluations then ranges from $2\binom{n}{2}$, if $G$ is *complete*, to $\sum_{i=1}^{n}(k+1)\binom{i}{k}\binom{i}{k+1}$, if $G$ is *edgeless* (for a complete graph $k > 1$ is not relevant). Some examples are given in tables 2 and 3.

|  | | | | | h | | | | | |
|---|---|---|---|---|---|---|---|---|---|---|
| n | a1 | a2 | a3 | a4 | a5 | a6 | a7 | a8 | a9 | a10 |
| 10 | 3.42 | 4.52 | 5.22 | 5.57 | 5.59 | 5.30 | 4.66 | 3.62 | 2.00 | - |
| 20 | 3.55 | 4.91 | 6.00 | 6.88 | 7.58 | 8.13 | 8.53 | 8.80 | 8.94 | 8.95 |
| 30 | 3.60 | 5.05 | 6.27 | 7.32 | 8.23 | 9.02 | 9.70 | 10.28 | 10.76 | 11.16 |
| 40 | 3.63 | 5.13 | 6.42 | 7.56 | 8.58 | 9.49 | 10.31 | 11.04 | 11.69 | 12.27 |
| 50 | 3.65 | 5.18 | 6.52 | 7.72 | 8.80 | 9.79 | 10.69 | 11.52 | 12.28 | 12.97 |
| 75 | 3.68 | 5.26 | 6.67 | 7.95 | 9.14 | 10.24 | 11.27 | 12.23 | 13.13 | 13.98 |
| 100 | 3.70 | 5.31 | 6.76 | 8.09 | 9.33 | 10.50 | 11.60 | 12.64 | 13.62 | 14.56 |
| 1000 | 3.80 | 5.54 | 7.18 | 8.74 | 10.25 | 11.71 | 13.12 | 14.50 | 15.85 | 17.17 |
| 10000 | 3.85 | 5.65 | 7.39 | 9.06 | 10.69 | 12.28 | 13.85 | 15.38 | 16.89 | 18.38 |
| 100000 | 3.88 | 5.72 | 7.51 | 9.25 | 10.95 | 12.63 | 14.28 | 15.91 | 17.52 | 19.11 |
| 1000000 | 3.90 | 5.77 | 7.59 | 9.37 | 11.13 | 12.86 | 14.57 | 16.26 | 17.93 | 19.59 |

Table 2. The number of evaluations $\tau$, in terms of $\log_n \tau$, that would have been made by *A* for *edgeless* graphs.

For an edgeless graph $A_{\mathbf{b}k}$ always makes the same number of evaluations as $A_{\mathbf{a}k}$, but also in general the numbers are more or less the same.

| n = 50 | | | | | | h | | | | | | |
|---|---|---|---|---|---|---|---|---|---|---|---|---|
| m | a1 | b1 | a2 | b2 | a3 | b3 | a4 | b4 | a5 | b5 | a6 | b6 |
| 100 | 3.49 | 3.48 | 4.97 | 4.96 | 6.20 | 6.19 | 7.24 | 7.24 | 8.13 | 8.12 | 8.87 | 8.86 |
| 200 | 3.38 | 3.36 | 4.80 | 4.77 | 5.91 | 5.89 | 6.79 | 6.77 | 7.46 | 7.45 | 7.94 | 7.93 |
| 300 | 3.30 | 3.26 | 4.63 | 4.59 | 5.62 | 5.60 | 6.32 | 6.30 | 6.76 | 6.75 | 6.96 | 6.95 |
| 400 | 3.17 | 3.16 | 4.43 | 4.42 | 5.26 | 5.26 | 5.73 | 5.73 | 5.86 | 5.86 | 5.64 | 5.64 |
| 500 | 3.10 | 3.09 | 4.26 | 4.25 | 4.93 | 4.93 | 5.18 | 5.18 | 5.03 | 5.03 | 4.47 | 4.47 |
| 600 | 3.03 | 3.03 | 4.07 | 4.07 | 4.57 | 4.57 | 4.58 | 4.58 | 4.14 | 4.14 | 3.25 | 3.25 |
| 700 | 2.94 | 2.94 | 3.85 | 3.85 | 4.12 | 4.12 | 3.85 | 3.85 | 3.10 | 3.10 | 1.84 | 1.84 |
| 800 | 2.85 | 2.85 | 3.61 | 3.61 | 3.63 | 3.63 | 3.00 | 3.00 | 0.00 | 0.00 | - | - |
| 900 | 2.73 | 2.74 | 3.29 | 3.29 | 2.94 | 2.94 | 1.77 | 1.77 | 0.00 | 0.00 | - | - |
| 1000 | 2.63 | 2.62 | 2.91 | 2.91 | 2.18 | 2.18 | 1.00 | 1.00 | 0.00 | 0.00 | - | - |
| 1100 | 2.44 | 2.44 | 2.24 | 2.24 | 0.00 | 0.00 | - | - | - | - | - | - |
| 1200 | 2.00 | 2.00 | 1.10 | 1.10 | 0.00 | 0.00 | - | - | - | - | - | - |

Table 3. The number of evaluations $\tau$, in terms of $\log_n \tau$, made by *A*, searching through 12 *random* graphs on 50 vertices and $m = (100, 200, …, 1200)$ edges.

The evaluations made by heuristic **b** are indeed more complex than the ones made by heuristic **a**, and this accounts for a substantial difference in runtime for certain combinations of *n*, *m* and *k*, but not to the extent that the workload is increased beyond polynomial time.

|   | Heuristic **a** | Heuristic **b** |
|---|---|---|
| 1. | Find $U = \overline{N}(S_i \cup \{v\})$ | Find $U = \overline{N}(S_i \cup \{v\})$ |
| 2. | *Nothing to do.* | Create $I \subset G$ such that $V(I) = U$. |
| 3. | Compute $|U|$. | Apply (1) to *I*. |

Let $w_{n\mathbf{a}}$ and $w_{n\mathbf{b}}$ be maximum amount of the work done by **a** and **b**, respectively, for a given *n*. The $w_{n\mathbf{b}}/w_{n\mathbf{a}}$ ratios for *n* = 100 are shown in Figure 1.

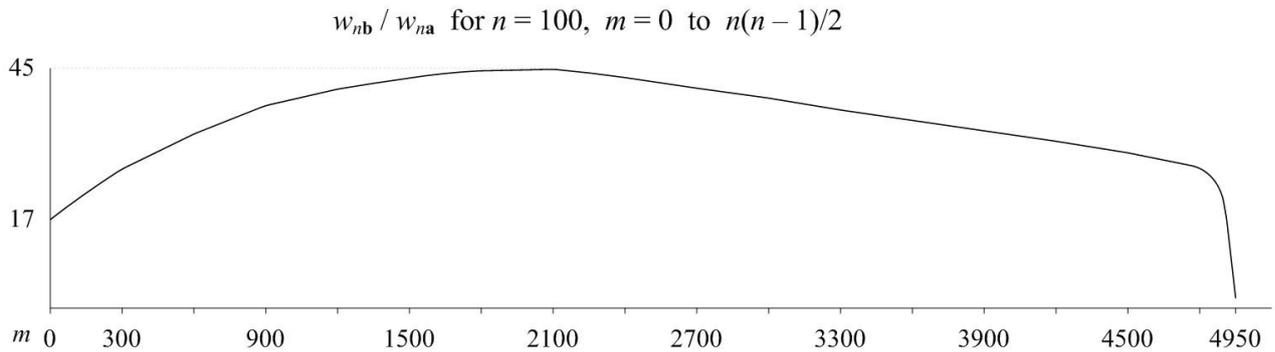

Figure 1.

The observed maxima of $w_{n\mathbf{b}}/w_{n\mathbf{a}}$ for *n* = 30, 40, …, 130 are

17, 21, 25, 31, 34, 39, 42, 45, 50, 54, 58.

If we write $w_{n\mathbf{b}} = Rnw_{n\mathbf{a}}$ the corresponding numbers for $R = w_{n\mathbf{b}}/nw_{n\mathbf{a}}$ are

0.57, 0.52, 0.50, 0.52, 0.49, 0.47, 0.45, 0.45, 0.45, 0.45.

The numbers are all put together in Table 4, which shows that for $n \geq 70$, $w_{n\mathbf{b}} < \frac{n}{2} w_{n\mathbf{a}}$.

| *n* | 30 | 40 | 50 | 60 | 70 | 80 | 90 | 100 | 110 | 120 | 130 |
|---|---|---|---|---|---|---|---|---|---|---|---|
| *max* $w_{n\mathbf{b}}/w_{n\mathbf{a}}$ | 17 | 21 | 25 | 31 | 34 | 39 | 42 | 45 | 50 | 54 | 58 |
| (*max* $w_{n\mathbf{b}}/w_{n\mathbf{a}}$)/*n* | 0.57 | 0.52 | 0.50 | 0.52 | 0.49 | 0.47 | 0.47 | 0.45 | 0.45 | 0.45 | 0.45 |

Table 4.